\documentclass{bmcart}

\usepackage[utf8]{inputenc} 

\def\includegraphics{}

\usepackage{moreverb}
\usepackage{lscape}
\usepackage{graphicx}
\usepackage{ctable}
\usepackage{subfig}
\usepackage{amsfonts}
\usepackage[T1]{fontenc}

\startlocaldefs
\endlocaldefs

\begin{document}

\begin{frontmatter}

\begin{fmbox}


\title{Application of Statistical Methods in Software Engineering: Theory and Practice}


\author[
   addressref={aff1,aff2},                   
   corref={aff1},                       
   email={tassio@tassio.eti.br}   
]{\inits{TS}\fnm{Tassio} \snm{Sirqueira}}
\author[
   addressref={aff1},
   email={}
]{\inits{MM}\fnm{Marcos} \snm{Miguel}}
\author[
   addressref={aff1},
   email={}
]{\inits{HD}\fnm{Humberto} \snm{Dalpra}}
\author[
   addressref={aff1,aff2, aff3},
   email={}
]{\inits{HD}\fnm{Marco Ant\^onio} \snm{Ara\'ujo}}
\author[
   addressref={aff1},
   email={}
]{\inits{HD}\fnm{Jos\'e Maria} \snm{David}}


\address[id=aff1]{
  \orgname{University Center Academia (UniAcademia)}, 
  \street{R. Halfeld, 1179 - Centro},                     %
  \postcode{36016-000}                                
  \city{Juiz de Fora},                              
  \cny{Brazil}                                    
}
\address[id=aff2]{%
  \orgname{Postgraduate Program in Computer Science - Federal University of Juiz de Fora (UFJF)},
  \city{Juiz de Fora},
  \cny{Brazil}
}
\address[id=aff3]{
  \orgname{Federal Institute of Education, Science and Technology
of Southeast Minas Gerais (IF Sudeste MG)}, 
  \street{Rua Bernardo Mascarenhas, 1283 - Bairro Fábrica},                     %
  \postcode{36080-001}                                
  \city{Juiz de Fora},                              
  \cny{Brazil}                                    
}


\begin{artnotes}
\end{artnotes}

\end{fmbox}


\begin{abstractbox}

\begin{abstract} 
The experimental evaluation of the methods and concepts covered in software engineering has been increasingly valued. This value indicates the constant search for new forms of assessment and validation of the results obtained in Software Engineering research. Results are validated in studies through evaluations, which in turn become increasingly stringent. As an alternative to aid in the verification of the results, that is, whether they are positive or negative, we suggest the use of statistical methods. This article presents some of the main statistical techniques available, as well as their use in carrying out the implementation of data analysis in experimental studies in Software Engineering. This paper presents a practical approach proving statistical techniques through a decision tree, which was created in order to facilitate the understanding of the appropriate statistical method for each data analysis situation. Actual data from the software projects were employed to demonstrate the use of these statistical methods. Although it is not the aim of this work, basic experimentation and statistics concepts will be presented, as well as a concrete indication of the applicability of these techniques.
\end{abstract}


\begin{keyword}
\kwd{Statistical Methods}
\kwd{Experimental Studies}
\kwd{Software Engineering}
\kwd{Experimental Evaluation}
\end{keyword}


\end{abstractbox}
%

\end{frontmatter}




\section{Introduction}\label{sec:introduction}
Software Engineering (SE) deals with the development, maintenance and management of high quality software systems in a cost effective and predictable manner. Research in  SE studies real world phenomena, addressing the development of new systems, modification of existing systems, as well as technology (process models, methods, techniques, tools or languages) to support SE activities. Also, it provides evaluations and comparisons of the effect of using technology to support complex interaction. Sciences that study real world phenomena, i.e. the empirical sciences, consist of collecting information based on observation and experimentation, instead of using logical or mathematical deductions \cite{sjoberg2007future}.
\\
An empirical approach of measuring SE technology, including industrial collaboration, started on a large scale in the 1970s from \cite{basili2002lessons} and \cite{boehm2005foundations} 's work. Currently, there is a growing appreciation of the experimental evaluation of methods and concepts covered in software engineering. This increase indicates the search for new forms of assessment and validation of results obtained in SE studies \cite{juristo2002reliable}. The production of significant empirical evaluations is facing many uncertainties and difficulties. Any researcher can forget or overlook a seemingly innocuous factor, which may prove to invalidate an entire work. Some essential experimental design guidelines can be ignored during the process, resulting in distrust regarding the validity of much of the work done \cite{miller1997statistical}.
\\
Studies in the computational area generally converge to build software, algorithms or models \cite{wainer2007metodos}. Results obtained from these studies are validated through evaluations, which are increasingly stringent. As an alternative to aid in the validation of either positive or negative results obtained from the study, we propose the use statistical methods. Statistics are applied in various areas of knowledge, providing methods for collection, organization, analysis and interpretation of data \cite{battisti2008metodos}. Statistical power is an inherent part of experimental studies, which employs significance tests, essential for planning, construction and validity of the findings of a study \cite{dybaa2006systematic}.
\\
Organizations usually question whether the expected results are being achieved \cite{rocha2012mediccao}. The answer to this question is not a trivial task, given that their actual performance is not always known \cite{florac1999measuring}. Statistical methods aim to provide numerical information, where data study and analysis  can be divided into three phases, \cite{da1999estatistica} (i) Data acquisition; (ii) Data description, classification and presentation; and (iii) Data conclusions. The author adds that the second phase is commonly known as Descriptive Statistics. The third phasis is called Inferential Statistics, which is one of the most important stages, since data collection and organization provides findings.
\\
This article consists of three other sections. In section 2, the theoretical assumptions are presented. In section 3, a practical approach for the use of the most common statistical methods is discussed. Finally, Section 4 describes the relevance of the application of statistical methods in studies geared towards Experimental Software Engineering.
\section[Theoretical Foundations]{Theoretical Foundations}
\begin{figure}[h!]
\setlength{\fboxsep}{0pt}%
\setlength{\fboxrule}{0pt}%
\centering
\includegraphics[keepaspectratio=true, height=150px]{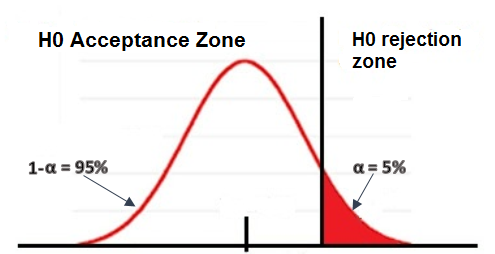}
\caption{Graphical representation of accepting the null hypothesis.}
\label{fig:1}
\end{figure}

\begin{figure}[!h]
\centering
\includegraphics[keepaspectratio=true, width=300px]{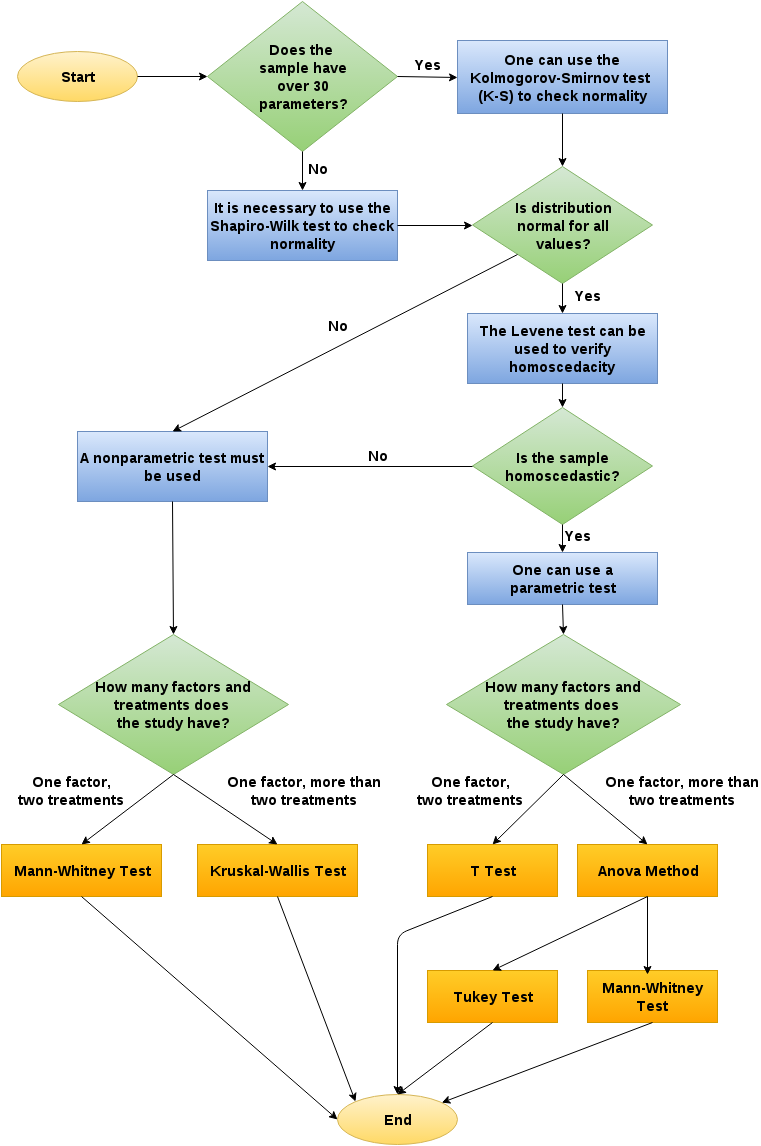}
\caption{Decision tree.}
\label{fig:2}
\end{figure}

The discussion on the role of statistical analysis in Experimental Software Engineering (ESE), indicates the underutilization of statistical power when discussing the results obtained \cite{miller1997statistical}. This fact leads to the appearance of failed research projects, as well as questionable validity of the results \cite{dybaa2006systematic}.
\\
Software Engineering commonly applies scientific methods in the evaluation of the benefits obtained from a new technique, theory or method related to software \cite{wohlin2012experimentation}. This method has traditionally been successfully applied to other sciences, especially social sciences. Social sciences resemble Software Engineering due to the increasing importance of the human factor in software, since it is rarely possible to establish laws of nature, as is done in physics or mathematics \cite{araujo2006metodos}.
\\
An experiment is an empirical examination and involves at least one treatment, a measurement of results, allocation units and some comparisons, from which the change can be inferred and assigned to the same \cite{sjoberg2007future}. According to \cite{araujo2006metodos}, the trial process is composed of four parts: definition, planning, execution and analysis. The statistical inference techniques can be applied to both the planning and the analysis of the trial. Planning formulates the hypothesis of the research, identifies the dependent (response) and independent (factors) variables, selects the participants and the methods of analysis. Furthermore, the study is projected, the instruments are defined and, finally, analyze the validity threats. The data analysis verifies the resulting graphs and descriptive statistics. It eliminates outliers (if applicable), analyzes data distributions and applies statistical methods in order to achieve the results.
\\
The purpose of an experimental study is to collect data in a controlled environment in order to confirm or dismiss the hypothesis. Hypothesis refers to a theory which seeks to explain a certain behavior of interest, to the research, and leads to the definition of independent or dependent variables. Independent variables represent the cause that affects the outcome of the trial process. The effect of the combination of the values of these variables refers to the dependent variables \cite{araujo2006metodos}. These variables can be quantitative, being expressed in numerical values that can be divided into interval and ratio scales, or qualitative, when they are not numeric and can be divided into nominal and ordinal scales \cite{battisti2008metodos}.
\\
According to \cite{hair2009analise}, only the nominal measurements indicate the type of data, and the only possible operation is to check wether the data has either one value or another, for example, the specification of the gender of individuals. Ordinal measures also give classes to the data, but you can sort them from highest to lowest and can cite as an example ``Level 2", which is smaller than ``Level 3" in the CMMI model. The interval measurement assigns a real number to the data, with the zero value being arbitrary in the scale. An example of interval measurement is the temperature in degrees Celsius. As for the ratio measurements, they assign a real number to the data, where zero is absolute,e.g., the distance in meters between two objects.
A comparison of the characteristics of these scales can be seen in Table \ref{tab:1}, as presented by \cite{araujo2006metodos}.
\begin{table}[h!]
\centering
\begin{tabular}{ccccc}
\hline 
Scale & Nominal & Ordinal & Interval & Ratio \\ 
\hline 
VALUE COUNTING & X & X & X & X \\ 
VALUE ORDINATION &  & X & X & X \\ 
EQUIDISTANT RANGES &  &  & X & X \\ 
VALUE ADDITION \\AND SUBTRACTION &  &  &  & X \\ 
VALUE DIVISION &  &  &  & X \\
\hline 
\end{tabular}
\caption{Characteristics of the values.}
\label{tab:1}
\end{table}
\\After collecting data from an experimental study, descriptive statistics are used to specify relevant characteristics, such as, (i) to indicate the middle of the set of observed data by means of central tendency measurements, (ii) to understand the average, median and mode values. The average is calculated from the sum of the collected values, divided by their number. The median is calculated by arranging the values in ascending (or descending) order and selecting the midpoint. The mode is calculated by counting the number of occurrences of each value and selecting the most common one. Other relevant measures are the minimum value, which is the lowest value among the data collected, the maximum value, which is the highest value among the data collected, the percentile, which divides the sample into values smaller than the size of sample quartile, which represents the 25\% percentile (or first quartile), the median (second quartile) and the 75\% percentile (third quartile) \cite{araujo2006metodos}.
\\
In order to measure the extent to which values are dispersed or concentrated in relation to their midpoint, dispersion measurements, including track, variance and standard deviation are used. Range represents the difference between the highest and the lowest value collected. Variance is the sum of the square difference between each value and the average of the collected values divided by the number of collected values, minus 1. Standard deviation is the root of the variance, which is one commonly used measurement \cite{araujo2006metodos}.
\\
A statistical hypothesis is a conjecture about unknown aspects in a data sample observed in a study, which can be proved or dismissed through a hypothesis test \cite{montgomery2010applied}. Hypothesis testing requires the specification of an acceptable level of statistical error, i.e. the risks the study is exposed to by decision-making \cite{neyman1928use}\cite{neyman1992problem}. To carry out a hypothesis test, a null hypothesis is defined, identified as H0, which is a statement according to which there is no difference between the parameter and the statistic you are comparing, and the alternative hypothesis, identified as H1, which contradicts H0 \cite{cooper2003metodos}. In general, it seeks to reject the null hypothesis in order to demonstrate that variations in the sample obtained with some intervention, or treatment, are not accidental.
\\
There are two possible types of error, such as, (i) the Type I error, which occurs when the statistical test indicates that there is a relation of cause and incorrect effect, also called False Positive, and (ii) Type II error, wherein the statistical test does not indicate the existence of a cause and effect relationship \cite{cooper2003metodos}.
\\
The probability of generating a Type I error (alpha) is related to the significance level of hypothesis testing. The lower the significance level is, the greater the assurance that a relationship is not identified \cite{shimakura2006estatistica}. The statistical significance of the result is an estimated measurement of the degree of accuracy of the result, i.e., the p-value is a decreasing index of the reliability of a result \cite{mahadevan2000probability}. In many research areas, the 0.05 p-level is customarily treated as an ``acceptable range" of error, since in particular research areas,such as in medical areas, the p-level can reach 0,001 and it is often called ``highly" significant, however, it is more susceptible to Type II error \cite{hair2009analise}.
\\
Figure \ref{fig:1} shows a graphical representation of accepting the null hypothesis, according to the 5\% significance level.
\\In order to reduce type I and II experimental errors, there are experimental designs which refer to how the treatment or factor levels are assigned to experimental units or portions \cite{markowitz1989mean}. Some of the main experimental designs are: completely randomized design, randomized block design and the Latin square design \cite{natrella2013experimental}.
\\
The completely randomized design is used when the variability between plots is small or nonexistent. In a randomized block design, the experimental material is divided into homogeneous groups. The objective at all stages of the experiment is to keep the error within each block, as small as possible in practical terms. For the Latin square design, treatments are grouped into repetitions in two distinct ways. This systematization of the blocks in two directions generically called ``lines" and ``columns" allows the elimination of variation effects in the experimental error.
\\
Hypothesis tests can be parametric, using the parameters of the distribution, or an estimate of them, or nonparametric \cite{siegel1956nonparametric}, which use posts assigned to the sorted data to calculate their statistics,they are free of the probability distribution of the data studied \cite{hollander2013nonparametric}. In parametric tests, it is assumed that their distribution is known. Although it is a better approach, it is necessary to ensure data normality and homoscedasticity, which refers to the less dispersed (concentrated) data around the regression line of the model. In nonparametric tests the sample distribution is not known, culminating in a less precise approach \cite{cooper2003metodos} \cite{camara2001estatistica}.
\\
When planning an experiment, one must choose the variable whose effects one wants to observe, which is called the ``factor". The categories of the factor under study are called ``treatments" \cite{natrella2013experimental}. In general, the purpose of an experimental study is to compare treatments, and verify whether these have the same effect on a measured characteristic, or whether at least one of them has a different effect when compared to the others \cite{natrella2013experimental}.
\\
Figure \ref{fig:2} shows a decision tree to facilitate the selection of an appropriate statistical method in order to carry out hypothesis tests, according to the characteristics of the sample and the study performed.
\\
Next you can see the concepts of the tests used in the flow chart as shown in Figure \ref{fig:2}, according to \cite{araujo2006metodos}.
\begin{itemize}
\item The Kolmogorov-Smirnov (KS) test enables the evaluation of the similarities between the distribution of two samples. It may also indicate the similarity in the distribution of a sample in relation to a classic distribution, such as the normal distribution.
\item The Shapiro-Wilk test is used to calculate the W value, which refers to the evaluation of a sample Xj with regard to normal distribution. In general, the relevant test is used for small data sets.
\item The Levene test considers the assumption that the variances are homogeneous if the W value is smaller than the value of the normal distribution.
\item The T or Student-t test is used to mean the comparison of two independent samples. In this, various tests are performed based on differences detected in the sample variances.
\item The ANOVA method is a statistical technique which aims to test the equality of the average of two or more groups.
\item The Tukey test, commonly used along with the ANOVA test, helps to identify samples whose averages diverge.
\item The Mann-Whitney test refers to an alternative, non-parametric test for the T test. In order to perform the Mann-Whitney test samples have to be independent, with ordinal, interval or ratio scales. 
\item The Kruskal-Wallis test is a nonparametric alternative to the analysis of variance (ANOVA) which, just like most parametric tests, is based on the replacement of values by their ranking in the set of all values.
\end{itemize}
\section[The decision tree in action]{The decision tree in action}
\begin{figure}[!h]
\centering
\includegraphics[keepaspectratio=true, width=350px]{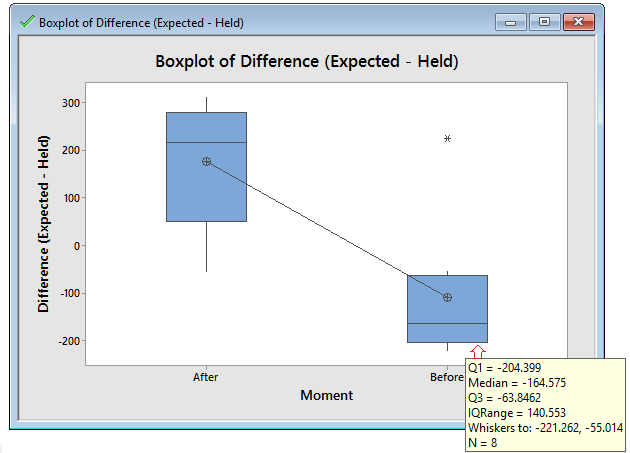}
\caption{Boxplot for the variables Difference (Expected - Held) x Moment.}
\label{fig:3}
\end{figure}

\begin{figure}[!h]
\centering
\includegraphics[keepaspectratio=true, width=350px]{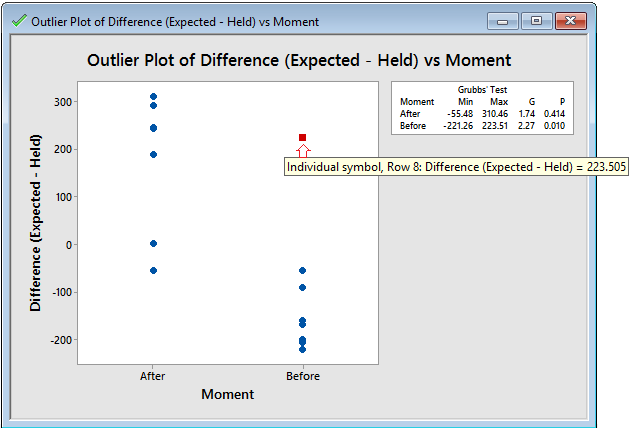}
\caption{Outlier analysis for the variables ``Difference (Expected - Held)" and ``Moment".}
\label{fig:4}
\end{figure}

This section presents an example whose main objective is to demonstrate the application of the aforementioned statistical techniques. It aims to demonstrate, step-by-step, how to use the methods laid out in the decision tree shown in Figure \ref{fig:2}. For the demonstration, data extracted from a real basis belonging to a software development company will be used, which is shown in Table \ref{tab:2}. It is desirable to verify the gains in the automated calculation of the development time, performed through a plugin, as opposed to the manual calculation of development time performed based on the experience of the developers involved in the project. The first 8 months (12/2013 to 07/2014) presented in the table reflect the planning period by the development team, the constant maintenance in the list of change orders. Values shown in the columns ``Expected Hours" and ``Held Hours" demonstrate the time spent by the staff, in hours, on the proper progress or completion of change requests. These are calculated manually, by empiricism, based on the experience of the developers involved. The following 8 months (08/2014 to 03/2015) comprise the period regarding the same of the maintenance planning, in which the development time is calculated through an automated calculation plugin. It was based on the history of maintenance performed by the team. The ``Difference (Expected - Held)" displays the time difference between expected and held time throughout the whole process.
\ctable[
caption={Planning Data (Expected and Held Values).},
label={tab:2},
botcap, 
sideways 
]
{ccccccc}
{
}
{
\hline
Year/Month & Held Hours & Expected Hours & Number of Cases & Cases Size & Difference (Expected - Held) & Moment \\ 
\hline
2013/12 & 259.878 & 100.000 & 36 & M & -159.878 & Before \\ 
2014/01	& 749.272 & 580.000	& 84 & L & -169.272	& Before \\
2014/02	& 570.343 & 480.000	& 74 & L & -90.343	& Before \\
2014/03	& 535.014 & 480.000	& 74 & L & -55.014	& Before \\
2014/04	& 311.262 & 90.000	& 33 & S & -221.262	& Before \\
2014/05	& 285.988 & 80.000	& 28 & S & -205.988	& Before \\
2014/06	& 279.633 & 80.000	& 28 & S & -199.633	& Before \\
2014/07	& 256.495 & 480.000	& 52 & M & 223.505	& Before \\
2014/08	& 437.427 & 680.000	& 52 & M & 242.573	& After \\
2014/09	& 450.845 & 395.367	& 58 & M & -55.478	& After \\
2014/10	& 225.472 & 517.222	& 75 & L & 291.750	& After \\
2014/11	& 602.305 & 791.996	& 95 & L & 189.691	& After \\
2014/12	& 450.147 & 452.305	& 62 & M & 2.158	& After \\
2015/01	& 327.089 & 516.024	& 70 & L & 188.935	& After \\
2015/02	& 258.536 & 503.461	& 65 & L & 244.925	& After \\
2015/03	& 310.315 & 620.772	& 80 & L & 310.457	& After \\
\hline
}
\\The Minitab tool \cite{minitab2000minitab}, version 17, was used in order to generate the analysis presented below.
\\
Table \ref{tab:3} shows the results of analyses of the data trend measurements which will be analyzed and discussed in the example presented in this article. These data were calculated based on Table \ref{tab:2}. The information presented in Table 3 can be generated in Minitab through the menu ``Stat -->Basic Statistics --> Display Descriptive Statistics" and selecting the variable ``Difference (Expected x Held)".
\begin{table}[h!]
\centering
\begin{tabular}{cc}
\hline 
TREND MEASUREMENTS &	VALUES - Difference (Expected - Held)\\
\hline 
AVERAGE	& 33.6\\
MEDIAN	& -26.4\\
MODE	& * (NUMBER OF MODES 0)\\
TRACK	& 531.7\\
MINIMUM	& -221.3\\
MAXIMUM	& 310.5\\
1st QUARTILE	& -166.9\\
3rd QUARTILES	& 237.8\\
VARIANCE	& 40244.0\\
STANDARD DEVIATION	& 200.6\\
\hline 
\end{tabular}
\caption{Analysis of measurements.}
\label{tab:3}
\end{table}
\\This information helps in analyzing the data generated in Figure \ref{fig:3}, as the value of ``Minimum", which is the smallest existing value for the variable ``Difference (Expected x Held)" and ``Maximum" being the highest value. The ``range" is the difference between the lowest and highest value in the variable ``Difference (Expected x Held)", the 1st and 3rd Quartiles are calculated based on the comparison with the variable ``Time" and the standard deviation is the difference between the median of each moment.
\\
The boxplot is a graph used to evaluate the distribution of the empirical data, which is formed by the first and third quartiles and the median. Analyzing the boxplot, we can observe the figures associated with the moments before and after the implementation of the plugin, in relation to the median drawn.
\\
The generation of this graph can be made in Minitab through the menu ``Graph  Box plot" option ``One Y / With Groups" by selecting the ``Difference (Expected x Held)" as the variable and, as for the category, by selecting the variable ``Moment".
\\
Another possible check on the presented data is the outlier analysis, which refers to the observations in the samples which are either very far from the others or are inconsistent when compared to them. These observations are also called abnormal, contaminant, strange, extreme or discrepant \cite{bamnett1994outliers}.
\\
It is important to know the reasons that lead to the appearance of outliers so that you can determine the correct way to treat them. Among the possible reasons for the appearance of outliers, measurement errors, data running or inherent variability of the elements of the population \cite{bamnett1994outliers} can be highlighted. An outlier resulting from collection or measurement errors should be discarded. However, if the observed value is possible, the outlier should not necessarily be discarded.
\\
In order to identify outliers it is necessary to calculate the median, lower quartile (Q1) and the upper quartile (Q3) of the samples. After that, the upper quartile must be subtracted from the lower quartile and the result must be stored in L). The values of the intervals Q3+1,5L and Q3+3L, and Q1-1,5L and Q1-3L will be considered outliers and can be accepted in the population. Values greater than Q3 + 3L and smaller than Q1-3L should be considered outliers. In this case, the origin of the dispersion must be investigated, because they are the most extreme points analyzed \cite{bamnett1994outliers}.
\\
Figure \ref{fig:4} shows the outliers regarding the analysis of the Difference (Expected - Held) variables. As can be seen, the samples did not have values that can be considered outliers, assuming a 5\% significance level.
\subsection[Analysis with 1 Factor and 2 Treatments using the parametric statistical method]{Analysis with 1 Factor and 2 Treatments using the parametric statistical method}
\begin{figure}[!h]
\centering
\subfloat[Moment: Before]{
\includegraphics[keepaspectratio=true, width=350px]{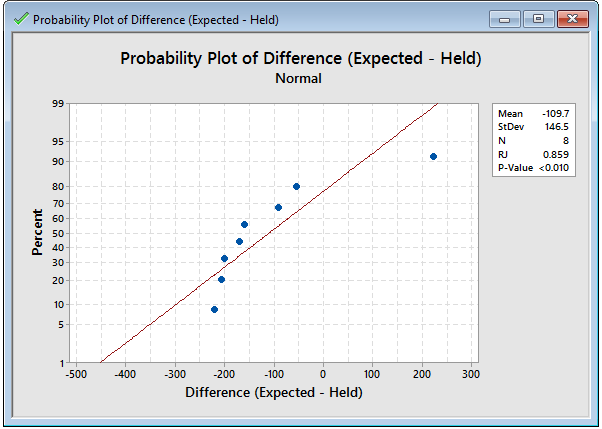}
\label{fig:5a}
}
\quad 
\subfloat[Moment: After]{
\includegraphics[keepaspectratio=true, width=350px]{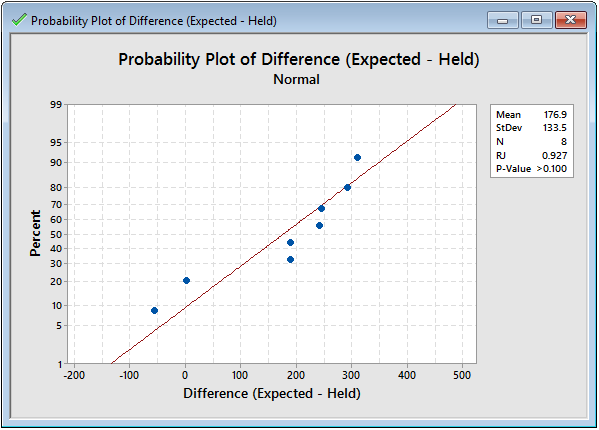}
\label{fig:5b}
}
\caption{Normality analysis for the variable Difference (Expected - Held).}
\label{fig:5}
\end{figure}

\begin{figure}[!h]
\centering
\includegraphics[keepaspectratio=true, width=350px]{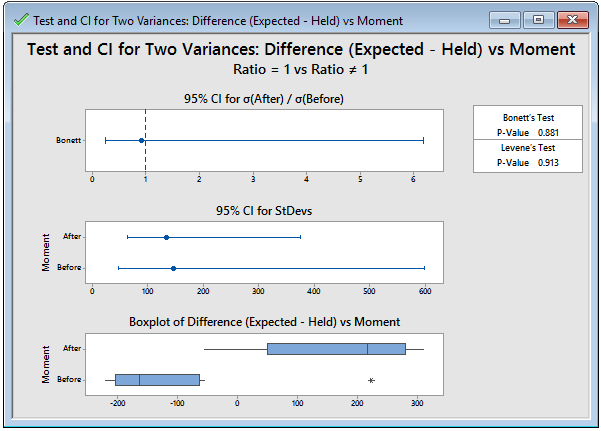}
\caption{Homoscedasticity analysis for the variables ``Difference (Expected - Held)" and ``Moment".}
\label{fig:6}
\end{figure}

\begin{figure}[!h]
\centering
\includegraphics[keepaspectratio=true, width=350px]{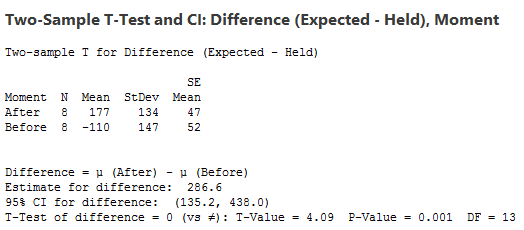}
\caption{Test T analysis for the variable ``Difference (Expected - Held)" x ``Moment".}
\label{fig:7}
\end{figure}

The analysis of Table \ref{tab:3} begins by checking data normality, for which the columns considered are ``Difference (Expected - Held)" and ``Moment". It can be seen that the table has fewer than 30 samples, thus the Shapiro-Wilk test is used, as shown in Figure \ref{fig:2}, which shows the decision tree. The following assumptions should be considered:
\begin{itemize}
\item H0 (null hypothesis): The samples have normal distribution.
\item H1 (alternative hypothesis): The samples did not show normal distribution.
\end{itemize}
To carry out this test in Minitab, the menu option ``Stat --> Basic Statistics --> Normality Test ..." and the variable ``Difference (Expected - Held)" must be selected. The normal test should be applied for each variable, individually. Since the variable ``Moment" is nominal and presents only two options (before and after), it does not require the normality test. In Figures \ref{fig:5a} and \ref{fig:5b} it is noted that, at a significance level of 5\%, samples are normal, since they have a p-value of 0.067, higher than the 0.05 significance level. It indicates the acceptance of H0, and samples show normal distribution.
\\Following the definition of the decision tree, we should check the homoscedasticity of the samples. For this purpose we need to check the following hypotheses:
\begin{itemize}
\item H0 (null hypothesis): Samples are homoscedastic.
\item H1 (alternative hypothesis): Samples are not homoscedastic.
\end{itemize}
Verification of the homoscedasticity test must be applied to the two variables involved in the hypothesis test. One must check whether the two samples are homoscedastic to each other. In Minitab, this can be done via the menu ``Start --> Basic Statistics --> 2 Variances ..." and by selecting the two variables to be compared.
\\
As shown in Figure \ref{fig:6}, which displays the Levene test to check the variance equality with a 5\% significance level, we obtain a p-value of 0.913. It appears that the samples are homoscedastic, given that the p-value is greater than the 5\% significance level, which indicates that a parametric test is needed.
\\For the statistical analysis, the parametric T test will be used according to the decision tree due to thethe number of treatments (2 factors: ``Before" and ``After" the adoption of time estimation plugin). This test is applied by considering the following assumptions:
\begin{itemize}
\item H0 (null hypothesis): There is no difference between the means.
\item H1 (alternative hypothesis): There are differences between the means.
\end{itemize}
In order to run T test in Minitab, we must access the menu ``Stat --> Basic Statistics --> 2-Sample t ...".
\\
Upon application of the respective test at a 5\% significance level, it is observed in Figure \ref{fig:7}, that with a p-value equal to 0.001, we can reject the null hypothesis that the means are statistically equivalent. Thus, there are significant differences, from a statistical point of view, for the samples.
\\
With the result of the analysis, we conclude that there is a significant difference between the means at a 5\%` significance level, with regard to time difference before and after adoption of the plugin. Thus, since the averages are significantly different, and the mean value is greater using the plugin, we can conclude that the use of the plugin was positive with regard to time difference.
\subsection[Analysis with 1 Factor and 2 Treatments using the non-parametric statistical method]{Analysis with 1 Factor and 2 Treatments using the non-parametric statistical method}
\begin{figure}[!h]
\centering
\subfloat[Time: Before]{
\includegraphics[keepaspectratio=true, width=350px]{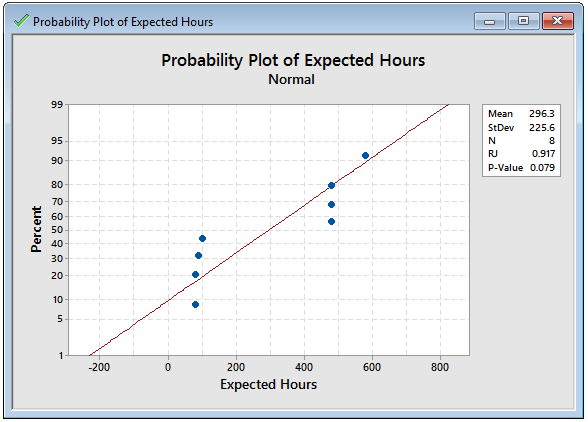}
\label{fig:8a}
}
\quad 
\subfloat[Time: After]{
\includegraphics[keepaspectratio=true, width=350px]{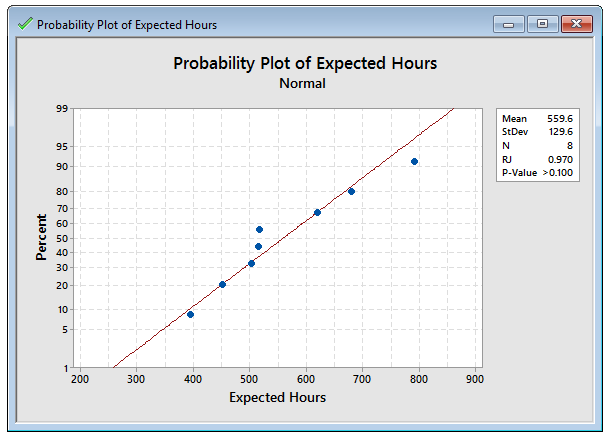}
\label{fig:8b}
}
\caption{Analysis of the variables ``Predicted Hours".}
\label{fig:8}
\end{figure}

\begin{figure}[!h]
\centering
\includegraphics[keepaspectratio=true, width=350px]{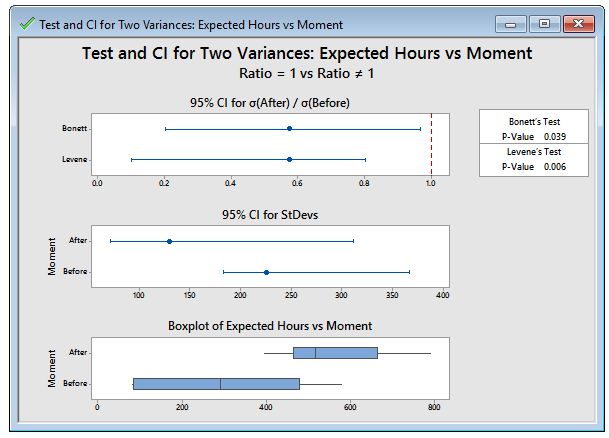}
\caption{Homoscedasticity analysis for the variables ``Expected Hours" and ``Moment".}
\label{fig:9}
\end{figure}

\begin{figure}[!h]
\centering
\includegraphics[keepaspectratio=true, width=350px]{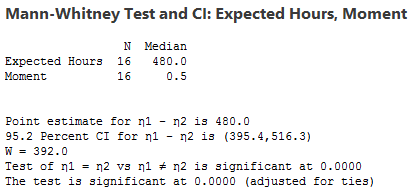}
\caption{Mann-Whitney test for the variable ``Difference (Expected - Held)" and ``Moment"
.}
\label{fig:10}
\end{figure}

Continuing the demonstration of statistical tests, the exemplification of the nonparametric tests is started. It starts with the normality check, for which the columns ``Expected Hours" and ``Moment" are taken into consideration. It can be seen that the table has less than 30 samples, therefore, the Shapiro-Wilk test is used, as shown in Figure \ref{fig:2}. 
\\The following assumptions should be considered:
\begin{itemize}
\item H0 (null hypothesis): The samples have normal distribution.
\item H1 (alternative hypothesis): The samples did not show normal distribution.
\end{itemize}
As shown in Figures \ref{fig:8a} (a) and \ref{fig:8b} (b), the samples exhibit a normal distribution with a p-value of 0.066 for the variable ``Predicted Hours". Therefore, their homoscedasticity must be checked. For this purpose, we need to check the following hypotheses:
\begin{itemize}
\item H0 (null hypothesis): Samples are homoscedastic.
\item H1 (alternative hypothesis): Samples are not homoscedastic.
\end{itemize}
As shown in Figure \ref{fig:9}, which displays the Levene test, comparing the variables ``Expected Hours" and ``Moment", a p-value of 0.006 is obtained. It is then noticed that the samples are not homoscedastic, given that one of the variables presented a p-value lower than the 5\% significance level, thus indicating the need for a non-parametric test.
\\
\\
In order to illustrate the use of a non-parametric method for one factor and two treatments, the Mann-Whitney test (alternative to the t-test) is used. To conduct this test, the variables ``Difference (Expected - Held)" and ``Moment" were compared, since the same have two treatments (``Before" and ``After" the adoption of the time estimation plugin). This test is applied considering the following assumptions:
\begin{itemize}
\item H0 (null hypothesis): There is no difference between the means.
\item  H1 (alternative hypothesis): There are differences between the means.
\end{itemize}
The non-parametric Mann-Whitney test can be performed by the Minitab menu ``Stat --> Nonparametrics --> Mann-Whitney ...".
\\ Figure \ref{fig:10} shows the Mann-Whitney test for the verification of average values, it appears that the samples do not have a significance level from a statistical standpoint, since they have an index lower than 5\%. Thus, the null hypothesis is rejected, which indicates that there is no significant difference between the means. The use of the plugin has brought significant benefits with regard to the difference in hours between the expected hours and the held hours in the project.
\\
After analyzing the time difference between the values before and after the adoption of the time estimation plugin, it can be said that there is a significant difference from a statistical point of view, considering a 5\% significance level, between the mean values observed.
\subsection[Analysis with 1 Factor and more than 2 Treatments using the parametric statistical method]{Analysis with 1 Factor and more than 2 Treatments using the parametric statistical method}
\begin{figure}[!h]
\centering
\includegraphics[keepaspectratio=true, width=350px]{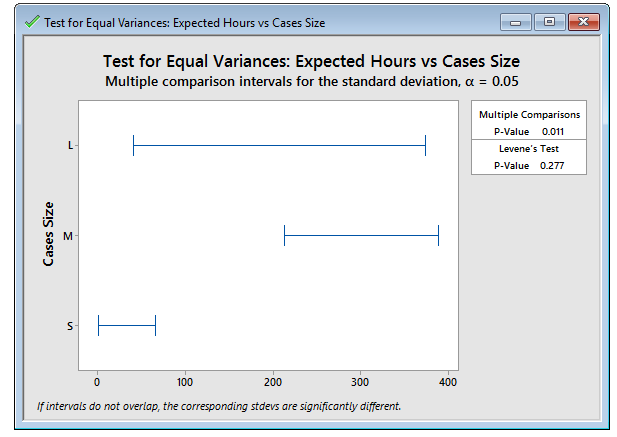}
\caption{Homoscedasticity test for the variable Expected x Held in hours.}
\label{fig:11}
\end{figure}

\begin{figure}[!h]
\centering
\includegraphics[keepaspectratio=true, width=350px]{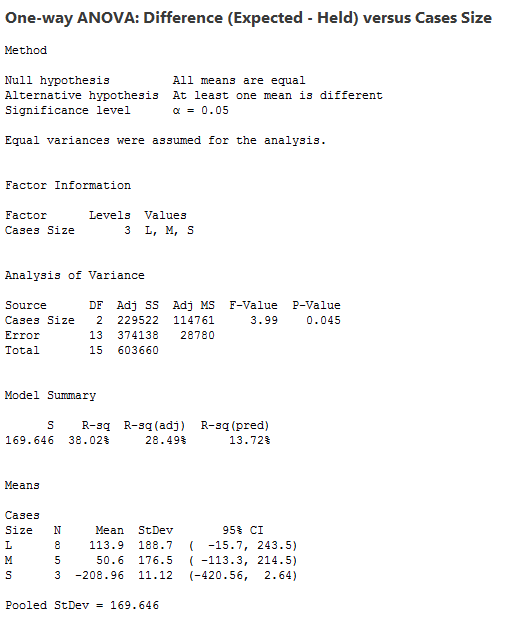}
\caption{Analysis of the ANOVA test for the variables ``Difference (Expected - Held)" and ``Cases Size".}
\label{fig:12}
\end{figure}

\begin{figure}[!h]
\centering
\includegraphics[keepaspectratio=true, width=350px]{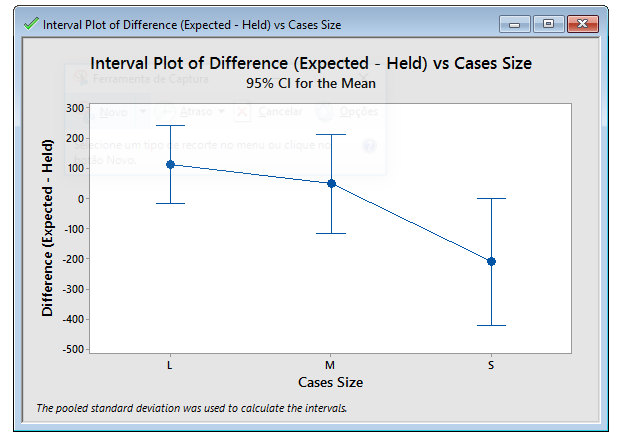}
\caption{Graphical analysis of the ANOVA for the variables ``Difference (Expected - Held)" and ``Cases Size".}
\label{fig:13}
\end{figure}

\begin{figure}[!h]
\centering
\includegraphics[keepaspectratio=true, width=350px]{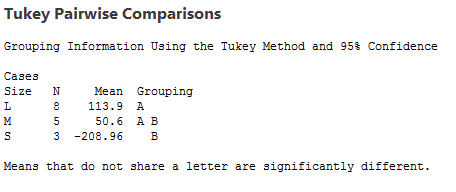}
\caption{Analysis of the ANOVA test using theTukey test for the variables ``Difference (Expected - Held)" and ``Cases Size".}
\label{fig:14}
\end{figure}

\begin{figure}[!h]
\centering
\includegraphics[keepaspectratio=true, width=350px]{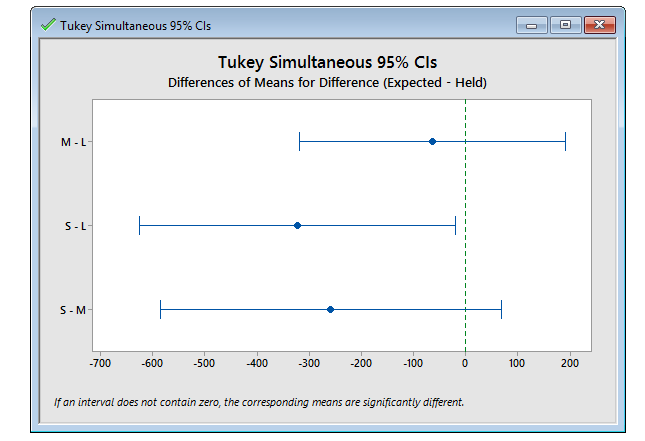}
\caption{Graphical analysis of the ANOVA test using the Tukey test for the variables ``Difference (Expected - Held)" and ``Cases Size".}
\label{fig:15}
\end{figure}

In order to illustrate the use of ANOVA, the normality of the variable ``Difference (Expected - Held)" contained in Figure \ref{fig:5}, and the separation of the variable ``Number of Cases" into three groups, which are, respectively,  ``Small" (S), where the number of cases is smaller than 33, ``Medium" (M), where the number of cases is between 34 and 66, and ``Large" (L), where the number of cases is higher than 66. This separation takes into account the fact that each version of the software developed by the company does not exceed 100 cases. This separation is represented in Table \ref{tab:2} in the ``Cases" column.
\\
For this purpose the following hypotheses needs to be considered:
\begin{itemize}
\item H0 (null hypothesis): Samples are homoscedastic.
\item H1 (alternative hypothesis): Samples are not homoscedastic.
\end{itemize}
As shown in Figure \ref{fig:11}, which shows the Levene test to check variance equality, with a 5\% significance level, we obtain a p-value of 0.220. It is then confirmed that the samples are homoscedastic, given that they present a p-value greater than the significance level of 5\%, thus indicating that a parametric test is needed.
\\
For the statistical analysis, the ANOVA parametric test (alternative to the nonparametric Kruskal-Wallis) will be used according to the decision tree, due to the comparison between the ``Difference (Expected - Held)" and ``Cases" of the samples. This test is applied considering the following assumptions:
\begin{itemize}
\item H0 (null hypothesis): There is no difference between the means.
\item H1 (alternative hypothesis): There are differences between the means.
\end{itemize}
The ANOVA test is available in Minitab through the menu ``Stat --> ANOVA --> One-Way ...'' by selecting ``Difference (Expected - Held)" as a variable, and ``Cases" as treatment.
\\
Upon the application of the respective test, at a 5\% significance level, and with a p-value equal to 0.045, Figure \ref{fig:12} shows that we can accept the null hypothesis that the mean values are statistically equivalent. Upon completion of this test, we can also get a graphical analysis of the ANOVA test, with a comparison of means according to the treatment, as shown in Figure \ref{fig:13}.
\\
By applying the Tukey test, available through the ``Comparisons ..." button in the ANOVA analysis, to the same variables discussed above, we can obtain the statistical analysis according to the treatment, which is used when the means analyzed by ANOVA are lower than the significance level, as shown both in Figure \ref{fig:14} and in the graphical analysis of Figure \ref{fig:15}.
\\
The ANOVA method shows that there is a difference between the expected and the held time in relation to the number of cases, and this analysis can be confirmed through the Tukey method, since the difference of mean values between a small number and a large number of cases is significant at a level of 5\%.
\subsection[Analysis with 1 Factor and 2 more Treatments using the non-parametric statistical method]{Analysis with 1 Factor and 2 more Treatments using the non-parametric statistical method}
\begin{figure}[!h]
\centering
\includegraphics[keepaspectratio=true, width=350px]{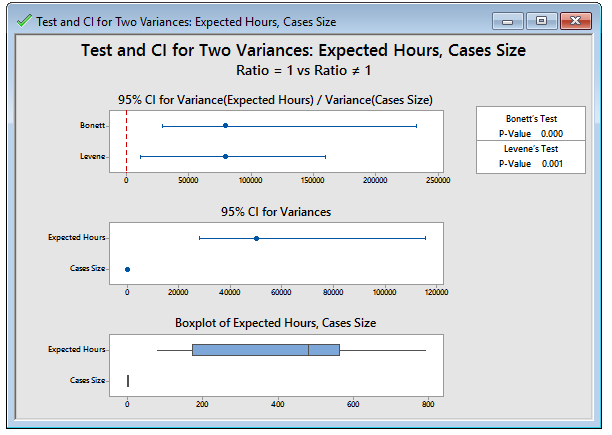}
\caption{Graphical analysis of the Levene test for the variables ``Expected Hours" and ``Cases Size".}
\label{fig:16}
\end{figure}

\begin{figure}[!h]
\centering
\includegraphics[keepaspectratio=true, width=250px]{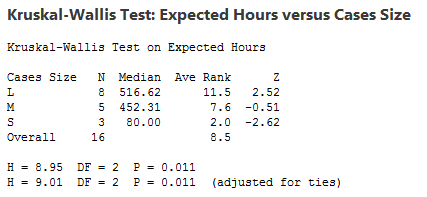}
\caption{Result of the Kruskal-Wallis test.}
\label{fig:17}
\end{figure}

\begin{figure}[!h]
\centering
\includegraphics[keepaspectratio=true, width=350px]{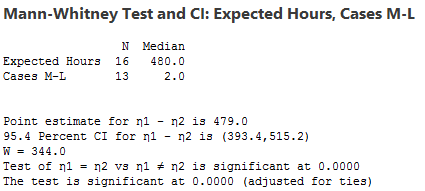}
\caption{Result of the Mann-Whitney test between ``Expected Hours" and Cases M-L.}
\label{fig:18}
\end{figure}

\begin{figure}[!h]
\centering
\includegraphics[keepaspectratio=true, width=350px]{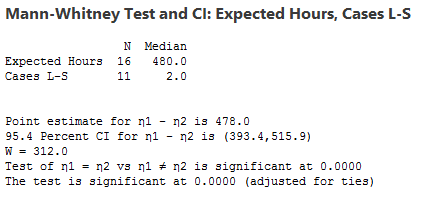}
\caption{Result of the Mann-Whitney test between ``Expected Hours" and Cases L-S.}
\label{fig:19}
\end{figure}

\begin{figure}[!h]
\centering
\includegraphics[keepaspectratio=true, width=350px]{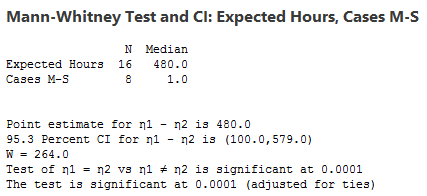}
\caption{Result of the Mann-Whitney test between ``Expected Hours" and Cases M-S.}
\label{fig:20}
\end{figure}

Considering the normality of the data arranged in Figure \ref{fig:8}, and that the decision tree is still taken into consideration, the next step refers to the comparison of mean values with regard to their homoscedasticity, which is applied to the variables ``Expected Hours", since normality has already been checked, as shown in Figure \ref{fig:8}, and to ``Cases Sizes", due to the number of treatments (3). For this purpose, the following assumptions need to be considered:
\begin{itemize}
\item H0 (null hypothesis): Samples are homoscedastic.
\item H1 (alternative hypothesis): Samples are not homoscedastic.
\end{itemize}
As shown in Figure \ref{fig:16}, which shows the Levene test to check variance equality , with a 5\% significance level, a p-value of 0.001 is obtained. It appears then that the samples are not homoscedastic, given that the p-value is lower than the 5\% significance level, thus indicating the need for a non-parametric test.
\\
We used the nonparametric Kruskal-Wallis test (a non-parametric alternative to ANOVA). This test is applied considering the following assumptions:
\begin{itemize}
\item H0 (null hypothesis): There is no difference between the means.
\item H1 (alternative hypothesis): There are differences between the means.
\end{itemize}
By applying this test at a 5\% significance level, a p-value equal to 0.011 can be seen in Figure \ref{fig:17}, thus indicating the rejection of the null hypothesis that the means are statistically equivalent.
\\
As the result obtained from the Kruskal-Wallis test was the rejection of the null hypothesis (H0), the Mann-Whitney test for comparison between groups must be applied. This second analysis demonstrates that there is a significant difference between the mean values, considering a 5\% significance level. The comparison between groups (M-L, L-S, M-S) can be seen in Figures \ref{fig:18}, \ref{fig:19} and \ref{fig:20}, respectively.
\\
As a result, there is a significant difference between the mean values at a 5\% significance level, which indicates the acceptance of the alternative hypothesis (H1).
\section[Conclusion]{Conclusion}
Based on the results obtained from the statistical analysis of the samples, we can mention that the use of the plugin showed improvements from a statistical point of view, compared to the empirical methodology initially adopted.
\\
The use of statistical methods has gained increasing attention within research in Experimental Software Engineering, thus showing their level of importance. It is also important to highlight that statistical methods are used for planning and conducting a study, data description and decision-making, where one can cite the hypothesis tests that are based on the risks associated with them. The formulation of hypotheses arising from statistical methods has been widespread. In order to decide whether a particular hypothesis is supported by a set of data, you must have an objective procedure by which to accept it or reject it. By formulating a decision on the null hypothesis (H0), two different errors can occur. The first, called a Type I error, comprises rejecting H0 when it is true. The second, called a Type II error, happens when one accepts H0 when it is false.
\\
In order to facilitate the choice of an appropriate statistical method depending on the sampling characteristics, this paper presents a decision tree, which shows the tests to be applied at each stage of data analysis.
\\
Statistical analyses of Samples may help researchers, software engineers and team leaders, in making important decisions about the risk associated with the project. This approach can produce greater maturity on the part of software engineers, considering that theywould no longer develop software based on assumptions, but based on facts, drawn from the statistical analyses of data from the software development processes.
\\
As a further work, we need to carry out additional experiments that consider the use of our proposal in other real world contexts.

\section{acknowledgements}
The authors thank IF Sudeste MG, CAPES and UFJF for the support and encouragement of research.



\end{document}